\documentclass[a4paper, amsfonts, amssymb, amsmath, reprint, showkeys, nofootinbib, twoside, superscriptaddress]{revtex4-2} 
\usepackage[english]{babel}
\usepackage[utf8]{inputenc}
\usepackage[colorinlistoftodos, color=green!40, prependcaption]{todonotes}
\usepackage{enumerate}
\usepackage{lipsum}
\usepackage{csquotes}
\usepackage[pdftex, pdftitle={Article}, pdfauthor={Author}]{hyperref}
\begin{document}
\title{GASP -- A Genetic Algorithm for State Preparation}

\author{Floyd M. Creevey}
\email{fcreevey@student.unimelb.edu.au}
\affiliation{School of Physics, University of Melbourne, VIC, Parkville, 3010, Australia.}
\author{Charles D. Hill}
\email{charles.hill1@unsw.edu.au}
\affiliation{School of Physics, University of Melbourne, VIC, Parkville, 3010, Australia.}
\affiliation{School of Mathematics and Statistics, University of Melbourne, VIC, Parkville, 3010, Australia.}
\author{Lloyd C. L. Hollenberg}
\email{lloydch@unimelb.edu.au}
\affiliation{School of Physics, University of Melbourne, VIC, Parkville, 3010, Australia.}

\date{February 17, 2023}

\begin{abstract}
    The efficient preparation of quantum states is an important step in the execution of many quantum algorithms. In the noisy intermediate-scale quantum (NISQ) computing era, this is a significant challenge given quantum resources are scarce and typically only low-depth quantum circuits can be implemented on physical devices. We present a genetic algorithm for state preparation (GASP) which generates relatively low-depth quantum circuits for initialising a quantum computer in a specified quantum state. The method uses a basis set of $R_x$, $R_y$, $R_z$, and CNOT gates and a genetic algorithm to systematically generate circuits to synthesize the target state to the required fidelity. GASP can produce more efficient circuits of a given accuracy with lower depth and gate counts than other methods. This variability of the required accuracy facilitates overall higher accuracy on implementation, as error accumulation in high-depth circuits can be avoided. We directly compare the method to the state initialisation technique based on an exact synthesis technique by implemented in IBM Qiskit simulated with noise and implemented on physical IBM Quantum devices. Results achieved by GASP outperform Qiskit's exact general circuit synthesis method on a variety of states such as Gaussian states and W-states, and consistently show the method reduces the number of gates required for the quantum circuits to generate these quantum states to the required accuracy.
\end{abstract}


\maketitle

\section{Introduction} \label{sec:Introduction}
    Quantum state preparation is key in many applications that require data input, such as finance \cite{black_pricing_1973, merton_theory_1973, orus_quantum_2019}, chemistry \cite{bauer_quantum_2020, cao_quantum_2019}, bioinformatics \cite{hollenberg_fast_2000}, machine learning \cite{heredge_quantum_2021}, and optimisation \cite{hassija_present_2020}. The ability to generate arbitrary quantum states efficiently and effectively, with high accuracy and low depth and gate count, will therefore be important in the future applications of quantum computing. Existing state preparation techniques \cite{niemann_logic_2016, shende_synthesis_2006} typically produce lengthy circuits with many qubit operations, or quantum gates, compromising their implementation on near-term hardware.  In this paper, we present a genetic algorithm for state preparation (GASP) which creates circuits for state preparation to a specified accuracy and depth in an evolutionary framework. This work builds on a relatively old idea on the application of genetic algorithms to quantum circuit evolution \cite{rubinstein_evolving_2001}.  For benchmarking purposes, of the methods that produce circuits for exact state preparation \cite{niemann_logic_2016, abdollahi_analysis_2006, daskin_decomposition_2013}, we focus here on the method by Shende, Bullock, and Markov (SBM) \cite{shende_synthesis_2006}, which has been encoded in the Qiskit library \cite{treinish_qiskitqiskit_2022}. The SBM approach produces an $n$-qubit arbitrary state using a circuit containing no more than $2^{n+1} - 2n$ CNOT gates. We benchmark the GASP method against Qiskit \cite{treinish_qiskitqiskit_2022} for Gaussian states and W-states in the context of simulation with gate noise and implementation on physical devices. Because the GASP circuits have much lower depth by design, the prospects for implementation on physical Noisy Intermediate Quantum (NISQ) devices are better, as we will show by explicit examples.
    
    The structure of the paper is as follows. Section \ref{sec:Genetic_Algorithms} will give a brief summary of genetic algorithms. Section \ref{sec:GASP} will describe the method presented in detail. Section \ref{sec:results} will present the results, and section \ref{sec:conclusion} our conclusions and potential future work.
\section{Genetic Algorithms} \label{sec:Genetic_Algorithms}
    A genetic algorithm is a classical optimisation technique that aims to mimic the process of biological evolution \cite{nielsen_quantum_2010}. The basic structure of a genetic algorithm is to first establish a population of individuals to evolve. This population can be many individuals or only a single individual. Generic genetic algorithms usually have many individuals and use crossover, however, if the population has only a single individual it is asexual, and only uses mutation. Each individual is an attempted solution to the given problem and is defined by a `chromosome' where each parameter in the chromosome represents a `gene'. This population is then subject to an iterative process where the individuals are selected by some selection criteria (rank selection, roulette wheel selection, etc.), and the selected individuals are bred together, and/or mutated. Each iteration of the genetic algorithm is referred to as a `generation'. For each generation, the fitness of each individual in the population is evaluated, by some objective fitness function, and the fittest individuals have the highest probability of being bred together, and/or mutated \cite{whitley_genetic_1994}. These individuals then form the next generation of the algorithm, in an effort to iteratively increase the maximum fitness. Generally, the algorithm is completed when a specific fitness is achieved, or when a given number of iterations are run without achieving a specific fitness. This method of optimisation allows solutions to be pulled out of local optima, and move towards global optima. The remainder of this section will outline the various aspects of a genetic algorithm and how these aspects may be applied to arbitrary quantum state synthesis.
    
    \textbf{\textit{Initial Population/Individual}}: In a sexual genetic algorithm the initial population is randomly generated with a specified population size parameter. Ideally, there is a considerable amount of divergence between different individuals, to allow for a broad area of the problem search space to be covered. In an asexual genetic algorithm, the initial population is only a single individual. Each individual in the population is instantiated with a chromosome defining a set of genes, the representation of these is subject to the problem being solved. Examples of possible chromosome representations include binary representations and list representations.
    
    \textbf{\textit{Crossover}}: Crossover is the process where two parents are bred together, producing a child that contains one-half of its genomic information from one parent, and one-half from the other. There are various methods of crossover in genetic algorithms, the main two being single-point crossover, and k-point crossover. The crossover point is where in the chromosome `list' the chromosome is split. Single-point crossover is where the crossover point is chosen and the child's genomic information is taken from one parent before the crossover point, and one parent after the crossover point. k-point crossover is a slightly more generalised version of this process, where there are k crossover points and the genomic information of the child ends up being $k+1$ sections alternating between parents \cite{spector_quantum_1999}.
    
    \textbf{\textit{Mutation}}: Mutation is the process where individual genes in an individual are randomly mutated, by some probability $p$, to increase genetic diversity in the population. In sexual genetic algorithms, the mutation rate is set low, as high mutation rates tend towards a primitive random search. In asexual genetic algorithms, the mutation rate is generally higher as it is the only means by which to introduce genetic diversity in the individual. The mutation operation allows the genetic algorithm to have increased genetic diversity, increasing the search space and allowing the algorithm to potentially escape local minima. 
    
    \textbf{\textit{Selection}}: Selection is the process by which the fittest individuals are chosen for crossover. The procedure generally involves assigning a probability of selection to each individual based on their fitness. There are many selection methods; common methods include roulette wheel selection, rank selection, and tournament selection. In \textit{roulette wheel selection} each individual is given a probability of being selected dependent on their fitness. This allows the fittest individuals to have the highest probability of passing genes on to the next generation. It also allows lucky unfit individuals to pass their genes on to the next generation, increasing genetic diversity. The advantage of this method is no genetic material is conserved. \textit{Rank selection} sorts the individuals by fitness and chooses the fittest individuals for crossover. The advantage of this method is you can more quickly converge to an optimal solution by virtue of only taking the fittest individuals. \textit{Tournament selection} randomly pairs individuals and selects the individual with the higher fitness of the two for crossover. This is an intermediary between the roulette wheel and rank selection, allowing potential faster convergence while maintaining a potentially higher level of genetic diversity.
\section{Genetic Algorithm for State Preparation (GASP)} \label{sec:GASP}
    \begin{figure*}
      \centering
        \includegraphics[width=\textwidth]{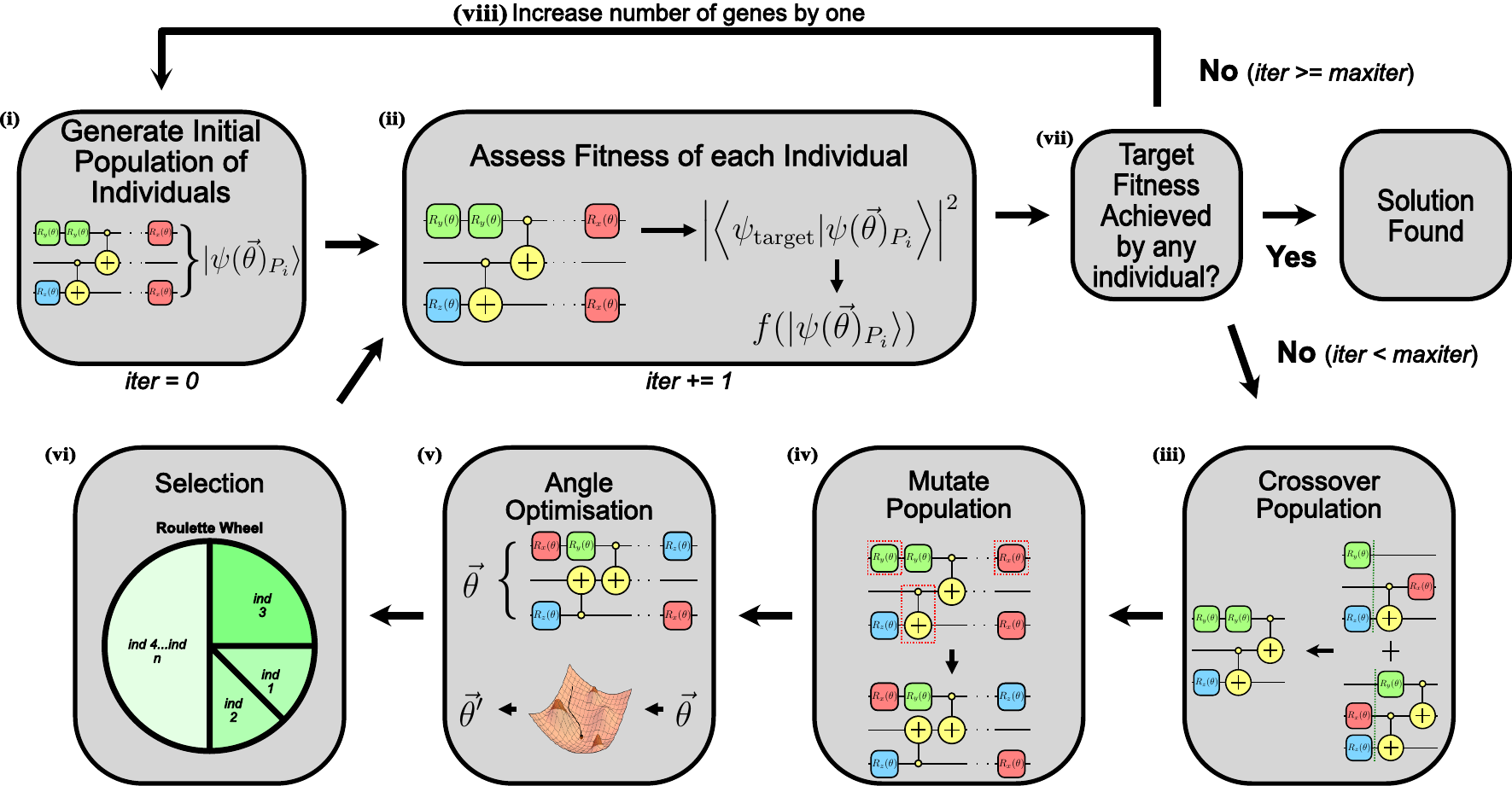}
        \caption{Overview of the GASP approach, given a desired target state vector $|\psi_{\rm{target}}\rangle$. \textbf{(\romannumeral 1)} Then, create initial population $\{P\}$, of individuals, $P_i$, which are each a quantum circuit, that generates the population states $|\psi(\vec{\theta})_{P}\rangle$, with the appropriate number of qubits and number of genes for the given state vector. \textbf{(\romannumeral 2)} Assess the fitness of the state vectors determined by each individual in the population: $f(|\psi(\vec{\theta})_{P_i}\rangle) = |\langle\psi_{\rm{target}}|\psi(\vec{\theta})_{P_i}\rangle|^2$. \textbf{(\romannumeral 3)} Apply crossover to the population, producing $|\psi(\vec{\theta})_{\rm{new}}\rangle$. \textbf{(\romannumeral 4)} Mutate the entire population with probability $p=5\%$. \textbf{(\romannumeral 5)} Run classical optimisation on each mutated individual to obtain the optimal $\theta$ values between $0$ and $2\pi$, to achieve the highest fitness for their generated circuit. \textbf{(\romannumeral 6)} Apply roulette wheel selection to the population, to select the individuals for the next generation based on their assessed fitness. \textbf{(\romannumeral 7)} Repeat until the desired fitness is achieved or $maxiter$ iterations since the last increase in fitness was achieved. \textbf{(\romannumeral 8)} If $maxiter$ iterations since the last increase in fitness, increase the number of genes by 1 and return to \textbf{(\romannumeral 1)}.}
        \label{GASP_flow}
    \end{figure*}

    In the quantum circuit space, the introduction of genetic algorithms has focused largely on producing circuits to generate specific unitaries \cite{spector_quantum_1999, lukac_evolving_2002, yabuki_genetic_2000, williams_quantum_1999, miranda_synthesis_2021, rubinstein_evolving_2001}. The flowchart describing GASP is depicted in Figure 1. In this work, the genetic algorithm is designed to evolve quantum circuits for the task of state preparation. In this context, an individual is a quantum circuit, and a gene is a single quantum logic operation (gate) in the given quantum circuit. A gene is represented,
    $$
    \rm{gene} = [\textit{q}_{\rm{t}}^\textit{i}, \textit{G}^\textit{i}, \textit{q}_{\rm{c}}^\textit{i}, \theta^\textit{i}],
    $$
    and an individual is represented,
    $$
    P_i = [\rm{gene}_1, \rm{gene}_2,\hdots, \rm{gene}_n],
    $$
    
    \noindent where each gene $(i = 0\ldots n)$ represents a gate application, $q_t$ is the target qubit, $G$ is the chosen gate type from $\{R_x, R_y, R_z, \rm{CNOT}\}$, $q_c$ is the control qubit if any, and $\theta$ is the rotation angle in the chosen gate. For a single qubit gate, $q_c$ is set to None. For a two-qubit gate $q_c$ is set to another qubit in the circuit, and $\theta$ is set to None, allowing the classical optimisation to only optimisation single-qubit rotation angles. The number of genes in an individual is dependent on how many gates the circuit contains. As such, a mutation in the genetic algorithm would be the changing of certain gates with a given probability.
    
    \textbf{\textit{Crossover}}: The crossover method used in GASP is a simple 1-point crossover with one half from each parent, applied to every generation (i.e.  crossover point is 50\%). In the quantum circuit context, this results in a new circuit containing half the gates from the first individual, and half the gates from the second individual.
    
    \textbf{\textit{Mutation}}: Mutation in GASP is applied every generation. The default probability for mutation is $5\%$, however, this parameter can be varied dependent on the problem to be solved. The basic mutation of a quantum circuit changes certain genes in the individual to other genes, resulting in different gates for the quantum circuit, changing the resultant state vector.
    
    \textbf{\textit{Fitness}}: Given the target state vector, $|\psi_{\rm{target}}\rangle$, and the resultant state vector of the current individual, $|\psi(\vec{\theta})_{P_i}\rangle$, GASP searches for the individual whose circuit produces the highest fitness. The fitness is calculated by the cost function, 
    $$
    f(|\psi(\vec{\theta})_{P_i}\rangle) = |\langle\psi_{\rm{target}}|\psi(\vec{\theta})_{P_i}\rangle|^2,
    $$
    the norm squared of the inner product between the target state vector and the individual's state vector, which is the similarity between the target state vector and the individual's state vector. The fitness of the individual will always be a value between 0 and 1. 
    
    \textbf{\textit{Selection}}: The method of selection used in GASP is roulette wheel selection. This allows genetic diversity to be maintained through generations of the algorithm, while also increasing fitness. This is done by initially summing the total fitness of the entire population of individuals, $f_T$, then giving each individual a normalised fitness relative to the fitness of the entire population, $p(|\psi(\vec{\theta})_{P}\rangle)$, i.e,
    
    \begin{eqnarray*}
    f_T = \sum_i f(|\psi(\vec{\theta})_{P_i}\rangle), \\ 
    p(|\psi(\vec{\theta})_{P_i}\rangle) = \frac{f(|\psi(\vec{\theta})_{P_i}\rangle)}{f_T}.
    \end{eqnarray*}
    
    \noindent The individuals in the next generation are then selected based on their respective fitnesses.
    
    \textbf{\textit{Algorithm}}: For a chosen $|\psi_{\rm{target}}\rangle$, a population of individuals is produced each with a certain number of `genes' based on the entanglement of the target state. The individual is the quantum circuit. Each `gene' is one of the four gates identified in the universal set listed in section \ref{sec:GASP}; $\{R_x, R_y, R_z, \rm{CNOT}\}$. The fitness of each individual is then assessed with the fitness function. Crossover is then applied to the population, to produce new individuals, doubling the population. Each individual in the population is then `mutated' at a probability of $5\%$. SLSQP optimisation \cite{kraft_software_1988} is then run on each individual in the population to find the optimal $\theta$'s for each given individual. The population is then subject to roulette wheel selection, to select the individuals for the next generation, halving the population back down to its original size. This process is then repeated, iteratively increasing the best fitness of the population. If the desired fitness is not achieved within \textit{maxiter} (1000 in the presented results) iterations of the last increase in fitness, the number of genes for each individual is increased by one, and the process restarts.  Once a desired target state vector $|\psi_{\rm{target}}\rangle$ (which also determines the number of qubits) is selected GASP can be broken down into the following steps:
    
    \begin{enumerate}[i.]
        \item The initial population $\{P\}$, of individuals, $P_i$, is created, that dictates the trial state vectors $|\psi(\vec{\theta})_{P_i}\rangle$, with the appropriate number of qubits and number of genes for the given state vector
        \item The fitness of the state vectors determined by each individual in the population: $f(|\psi(\vec{\theta})_{P_i}\rangle) = |\langle\psi_{\rm{target}}|\psi(\vec{\theta})_{P_i}\rangle|^2$ is assessed
        \item Crossover is applied to the population, generating $|\psi(\vec{\theta})_{\rm{new}}\rangle$
        \item The entire population is mutated with probability $p=5\%$.
        \item Classical optimisation is run on each mutated individual to obtain the optimal $\vec{\theta}$ values between $0$ and $2\pi$, which achieve the highest fitness for their generated circuit
        \item Roulette wheel selection is applied to the population, to select the individuals for the next generation based on their assessed fitness.
        \item Steps \romannumeral2-\romannumeral6 \space are repeated until the desired fitness is achieved or $maxiter$ (where $maxiter$ is a parameter set prior to the start of the algorithm) iterations since the last increase in fitness was achieved
        \item If $maxiter$ iterations since the last increase in fitness are achieved, increase the number of genes by 1 and return to step \romannumeral 1
    \end{enumerate}
    
    \noindent In GASP, the genetic algorithm is being utilised to determine the course-grained optimisation of quantum circuit structure, while a much better fine-grained algorithm, SLSQP, was used for determining the optimal angles for each circuit structure generated. This allows the optimal, or at least close to optimal, fitness for each generated circuit structure to be achieved.
\section{Results - GASP vs. Qiskit} \label{sec:results}

    \subsection{Gaussian States}
    
\begin{figure*} 
      \centering
        \includegraphics[width=\textwidth]{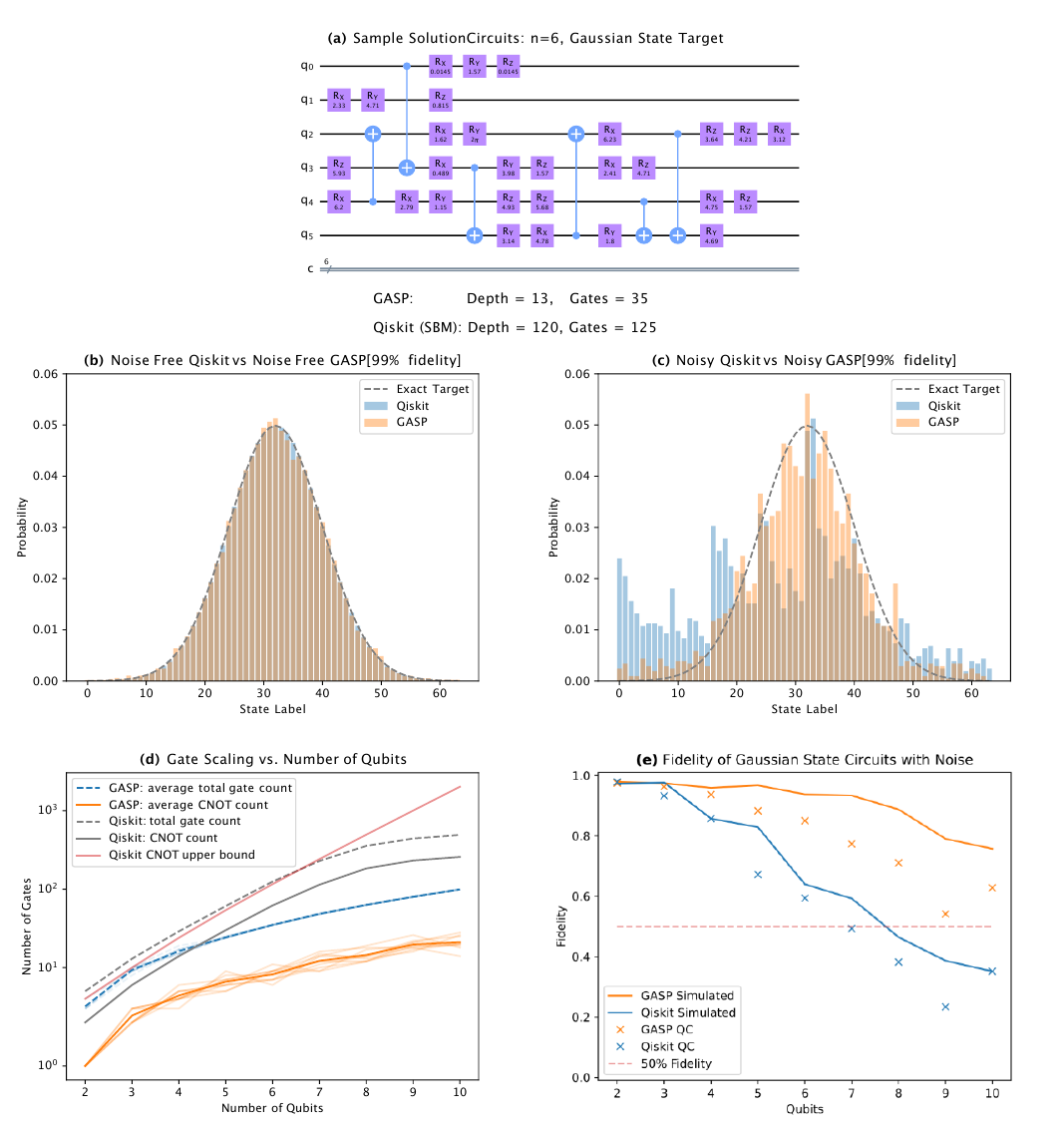}
        \caption{\textbf{(a)} Comparison between sample solution circuits generated by GASP and Qiskit for a 6 qubit Gaussian. GASP produced a circuit with a depth of 13 and 35 gates, Qiskit produced a circuit with a depth of 120 and 125 gates. \textbf{Note:} these are circuits before compilation on real hardware.  \textbf{(b)} Comparison between Qiskit and GASP(99\% fidelity) resultant distributions with no noise for a 6 qubit Gaussian state. The grey dashed line is the exact distribution, the orange bars are GASP, and the blue bars are Qiskit. \textbf{(c)} Comparison between Qiskit and GASP(99\% fidelity) resultant distributions in the presence of noise for a 6 qubit Gaussian state. The noise model is that of IBM's \textit{ibmq\_guadalupe} machine. 16384 shots were used. The grey dashed line is the exact distribution without noise, the orange bars are GASP, and the blue bars are Qiskit. \textbf{(d)} Gate comparison between the Qiskit and GASP(99\% fidelity) for Gaussian states as the number of qubits varied from 2 to 10. \textbf{(e)} Comparison between GASP(99\% fidelity) and Qiskit for Gaussian states as the number of qubits varied from 2 to 10. Lines represent simulations with noise, crosses represent results from IBM's \textit{ibmq\_guadalupe} machine. 16384 shots were used. The noise model is that of IBM's \textit{ibmq\_guadalupe} machine.}
        \label{fig:gaussian_results}
    \end{figure*}

    Gaussian states have relatively low entanglement and are defined as,
    $$
    |\psi_G\rangle = \frac{1}{\sqrt{2^n}} \sum_{i=0}^{2^n-1}g(x)|x\rangle,
    $$
    where $g(x) = \frac{1}{\sigma\sqrt{2\pi}}\exp{(-\frac{1}{2}\frac{(x - \mu)^2}{\sigma^2})}$, $\mu$ is the mean, and $\sigma$ is the standard deviation of the desired Gaussian. For the purposes of this paper, we let $\mu = \frac{2^n}{2}$, and $\sigma = \frac{2^n}{8}$. In Figure \ref{fig:gaussian_results}(a) we show an example GASP circuit, comparing the depth and gate count with that produced by Qiskit's initialise function. A comparison of the states produced by Qiskit's initialise function, and GASP, in the absence of noise, is shown in Figure \ref{fig:gaussian_results}(b) for a 6 qubit Gaussian state. In the zero-noise regime, the data shows that the Qiskit method produces the target exactly as expected, and the GASP is within the specified fidelity tolerance (99\% in this case). For the same 6 qubit Gaussian, a comparison of the states produced by Qiskit's initialise function, and GASP simulated, in the presence of noise (modelled from \textit{ibmq\_guadalupe}) is shown in Figure \ref{fig:gaussian_results}(c), 16384 shots were used. It can be seen that GASP performs much better in reproducing the desired target state vector in the presence of noise, though not perfectly. This shows that a slight reduction in the accuracy of the circuit, for a large reduction in circuit depth improves the desired outcome state in the presence of noise. GASP producing circuits shorter by orders of magnitude allow more realistic circuits to be implemented on NISQ-era hardware. Figure \ref{fig:gaussian_results}(d) shows the comparison between GASP and Qiskit for Gaussian states, in terms of gate scaling vs. the number of qubits. Figure \ref{fig:gaussian_results}(e) shows the improved performance of the circuits generated by GASP relative to Qiskit when simulated with noise, and run on IBM's \textit{ibmq\_guadalupe} machine, averaged over 10 tests as the number of qubits varied from 2 to 10 (noting that this measure does not include the phase information).
    
    \subsection{W-states}
    
\begin{figure*}
      \centering
        \includegraphics[width=\textwidth]{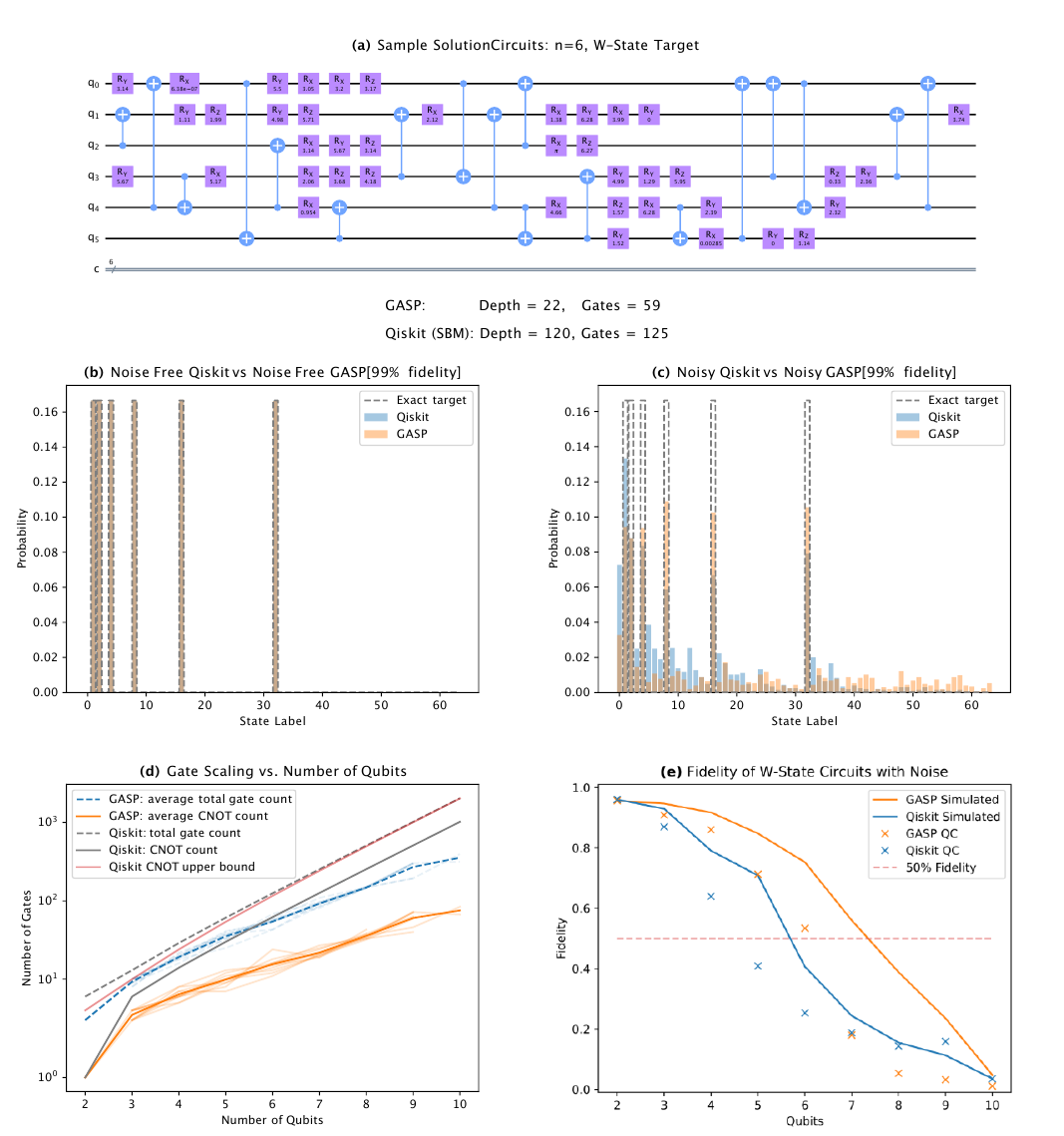}
        \caption{\textbf{(a)} Comparison between sample solution circuits generated by GASP and Qiskit for a 6 qubit W state. GASP produced a circuit with a depth of 22 and 59 gates, and Qiskit produced a circuit with a depth of 120 and 125 gates. \textbf{Note:} these are circuits before compilation on real hardware. \textbf{(b)} Comparison between Qiskit and GASP(99\% fidelity) resultant distributions with no noise for a 6 qubit W state. The grey dashed line is the exact distribution, the orange bars are GASP, and the blue bars are Qiskit. \textbf{(c)} Comparison between Qiskit and GASP(99\% fidelity) resultant distributions in the presence of noise for a 6 qubit W state. The noise model is that of IBM's \textit{ibmq\_guadalupe} machine. 16384 shots were used. The grey dashed line is the exact distribution without noise, the orange bars are GASP, and the blue bars are Qiskit. \textbf{(d)} Comparison between the Qiskit and GASP(99\% fidelity) for W-states as the number of qubits varied from 2 to 10. \textbf{(e)} Comparison between GASP(99\% fidelity) and Qiskit for W-states as the number of qubits varied from 2 to 10. Lines represent simulations with noise, crosses represent results from IBM's \textit{ibmq\_guadalupe} machine. 16384 shots were used. The noise model is that of IBM's \textit{ibmq\_guadalupe} machine.}
        \label{fig:w_results}
    \end{figure*}

    W-states generally have higher entanglement and are defined as,
    $$
    |\psi_W\rangle = \frac{1}{\sqrt{n}}(|100...0\rangle + |010...0\rangle + ... + |000...1\rangle),
    $$
    where $n$ is the number of qubits. A comparison of the produced states by Qiskit's initialise function, and GASP, with no noise, is shown in Figure \ref{fig:w_results}(b). Figure \ref{fig:gaussian_results}(c) shows the comparison between Qiskit and GASP(99\% fidelity) resultant distributions in the presence of noise for a 6 qubit W state. The noise model is that of IBM's \textit{ibmq\_guadalupe} machine. 16384 shots were used. The grey dashed line is the exact distribution without noise, the orange bars are GASP, and the blue bars are Qiskit. Figure \ref{fig:gaussian_results}(d) shows the comparison between the Qiskit and GASP(99\% fidelity) for W-states as the number of qubits varied from 2 to 10. Figure \ref{fig:gaussian_results}(e) shows the performance of the circuits generated by GASP and Qiskit when simulated with noise, averaged over 10 tests as the number of qubits varied from 2 to 10 noting that this measure does not include the phase information.
    
    As can be seen in the results shown here, GASP consistently outperforms Qiskit in the number of total gates and the number of CNOT gates required; by more than two orders of magnitude in the higher qubit tests. It seems that GASP is a lower polynomial complexity in the number of gates required compared to Qiskit's initialisation method. It should be noted that the number of gates required for the Gaussian state generation is fewer than that of the W-states. This is likely due to Gaussian states being less entangled than W-states. It can be seen that the circuits produced by GASP have higher noise robustness than circuits produced by Qiskit's initialisation. However, the length of circuits produced by both techniques at high numbers of qubits have so many gates that the noise overwhelms the ability of the circuit to produce the desired state. It should also be noted that at useful fidelities (those above 50\%), GASP outperforms Qiskit's initialisation.
\section{Conclusion} \label{sec:conclusion}
    In this paper, we have proposed and demonstrated a state preparation method based on a genetic evolutionary approach. Benchmarking GASP against Qiskit's initialisation method, the results show that in the noisy regime relevant to implementation on actual hardware, GASP significantly outperforms the exact approach through superior circuit compression, by more than an order of magnitude. As GASP is a stochastic algorithm, there is an increase in run time over the deterministic algorithms and an introduced uncertainty in the ability to produce a solution. However, the significant reduction in both the total gate count and the number of required CNOT gates may outweigh these for the application of GASP to the initialisation of quantum states in circuit lengths feasible on NISQ-era hardware.
    
    Note: During the preparation of this work, a recent paper by Rindell et. al \cite{rindell_generating_2022} studying state preparation using a genetic algorithm was posted on the arXiv. Their method is similar to the GASP approach presented here, though differs in that they used a Fast Non-dominated Sorting Genetic Algorithm implemented in the DEAP Python package \cite{fortin_deap_2012}, and a different gate set of $\{R_z(\theta), X, \sqrt{X}, \rm{CNOT\}}$. They also do not use classical optimisation, instead opting to adjust $\theta$ by adding a value from a selected Gaussian distribution. Their method was applied to the production of Haar random states up to 5 qubits, demonstrating similar fidelity improvements in the presence of noise. 
\section*{Acknowledgements} \label{sec:acknowledgements}
This research was supported by the University of Melbourne through the establishment of the IBM Quantum Network Hub at the University. FMC is supported by an Australian Government Research Training Program Scholarship. This research was supported by The University of Melbourne’s Research Computing Services and the Petascale Campus Initiative.


%


\end{document}